\documentclass[aps,twocolumn,superscriptaddress,notitlepage,longbibliography]{revtex4-1}
\usepackage{amsmath,amssymb,bm,bbm,mathrsfs,braket,color,graphicx,float,comment,amsfonts,dsfont}
\usepackage[colorlinks,citecolor=blue,urlcolor=blue]{hyperref}

\DeclareMathAlphabet\mathbfcal{OMS}{cmsy}{b}{n}

\usepackage{extarrows} 
\usepackage{mathrsfs}

\definecolor{purple}{rgb}{0.7, 0., 0.8}
\definecolor{grun}{rgb}{0.0, 0.7, 0.0}

\usepackage{tikz,lipsum,lmodern}
\usepackage[most]{tcolorbox}

\begin{document}

\title{Spatial Engineering of Topological Quantum Matter with Structured Light}

\author{Matias Mustonen}
\affiliation{Photonics Laboratory, Physics Unit, Faculty of Engineering and
Natural Sciences, Tampere University, FI-33014 Tampere, Finland}
\affiliation{Computational Physics Laboratory, Physics Unit, Faculty of Engineering and
Natural Sciences, Tampere University, FI-33014 Tampere, Finland}

\author{Ali G. Moghaddam} \email{ali.moghadam@aalto.fi}
\affiliation{Department of Applied Physics, Aalto University, FI-00076 Espoo, Finland}
\affiliation{Computational Physics Laboratory, Physics Unit, Faculty of Engineering and
Natural Sciences, Tampere University, FI-33014 Tampere, Finland}

\author{Teemu Ojanen}
\affiliation{Computational Physics Laboratory, Physics Unit, Faculty of Engineering and
Natural Sciences, Tampere University, FI-33014 Tampere, Finland}

\author{Marco Ornigotti}\email{marco.ornigotti@tuni.fi}
\affiliation{Photonics Laboratory, Physics Unit, Faculty of Engineering and
Natural Sciences, Tampere University, FI-33014 Tampere, Finland}

\begin{abstract}
    Structured light provides a powerful means of controlling matter while itself exhibiting nontrivial spatial topology. Here, we show that these spatial structures can be harnessed to locally engineer the topological properties of matter. In a paradigmatic example of a topological insulator with Dirac surface states driven by an optical skyrmion beam, we show that structured light spatially tailors the effective Hamiltonian, yielding position-dependent topological gaps and Dirac velocities. The resulting gap modulation can create topological and trivial regions separated by optically defined interfaces, without introducing material boundaries or chemical modifications. We further show that the geometry of these interfaces can be controlled through the spatial structure of the light, in particular its orbital angular momentum. Our results establish structured light as a route to dynamically pattern topological phases and open a broader perspective on the interplay between topology in light and matter, including the possibility of transferring topological characteristics between the two.
\end{abstract}

\maketitle

\emph{Introduction}.---The interaction between light and matter has long been a cornerstone of physics, leading to fundamental insights and technologies like lasers and advanced spectroscopy \cite{Cohen-Tannoudji1998,rubio2018lightmatter,nori2019lightmatter,Ultrastrong2019lightmatter,rivera2020lightmatter,gutzler2021lightmatter,frisk_kockum_ultrastrong_2019,lininger2023chirality,hubener_quantum_2024,Stammer_PRXQuantum,svendsen_combining_2021,hubener_engineering_2021,goto_chern_2025,appugliese_breakdown_2022}. Intense light fields not only probe materials but also actively control and reshape their intrinsic properties. Optical driving can dynamically tailor electronic band structures and elementary excitations \cite{sentef2024floquet}, and theoretical studies suggest it can even manipulate strongly correlated electron phenomena \cite{bloch2022strongly}, potentially inducing or enhancing non-equilibrium correlated phases, including light-induced superconductivity \cite{mitrano2016superconductivity,demsar2016light}. \emph{Floquet engineering}, using time-periodic driving, has emerged as a powerful technique to create new quantum states, particularly effective in producing non-equilibrium topological phases with unique properties not found in static systems \cite{Aoki2009floquet,Demler2011floquet,lindner2011floquet,rechtsman2013floquet,
Dehghani2015floquet,mciver2020floquet,merboldt2024floquet,floquet_topo_exp_2026,oka_floquet_2019,Kitagawa2010,Rudner_2015}.

On a seemingly parallel trail, generating structured light beams. i.e., light fields with customizable properties across all of their degrees of freedom, is nowadays the standard for complex light fields in many different fields of fundamental and applied physics \cite{forbesBook}. Examples include beams with orbital angular momentum (OAM) \cite{ref1}, polarization singularities \cite{Dennis_Singular_optics_2009}, vector beams with nonuniform polarization pattern \cite{Rosales-Guzman_2018}, and complex topological light structures, such as optical skyrmions \cite{ref4,skyrmionBeams,du2019deep,he2024optical,stoyanov2023optical}.

\begin{figure}[!]
\centering
\includegraphics[width=0.8\linewidth]{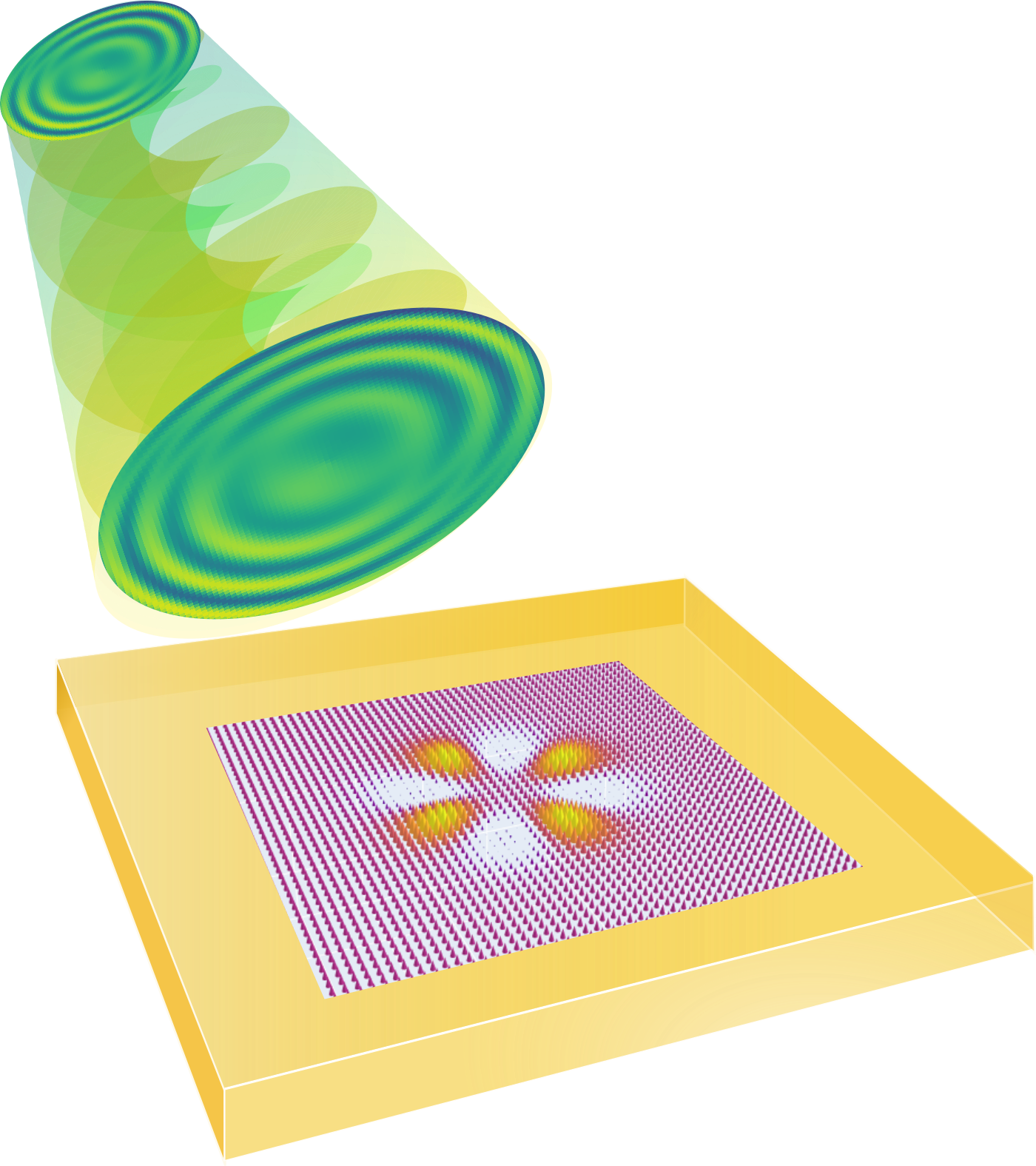}
\caption{ Pictorial representation of the topological light-matter interaction scheme described in this work. A structured light beam (green) impinges onto a topological insulator (yellow). As a result of the interaction, electron spins acquire a nontrivial, spatially inhomogeneous texture, imprinted by the impinging light beam.
}
\label{fig1} 
\end{figure}

Despite the vast and detailed analysis in both Floquet engineering and structured light, the interaction between Floquet-driven topological matter and topological light fields remains largely unexplored, and are limited to the interaction of OAM beams with topological insulators (TIs) in the presence of disorder \cite{OAMvsTI}, or Chern insulators created through the interaction of circularly polarized plane waves with graphene-like materials \cite{ChernCircular}. To the best of our knowledge, TIs driven by light carrying nontrivial topological features, like an optical skyrmion, has not been yet investigated.

Motivated by this, in this Letter we theoretically investigate the use of structured light to spatially engineer the topological properties of matter (see Fig.~\ref{fig1}). Considering a Dirac-like topological insulator driven by a monochromatic optical skyrmion within the Floquet formalism, we show that light can spatially modulate both the topological gap and the effective Dirac velocities. In particular, a local sign reversal of the gap creates topological and trivial regions separated by optically defined interfaces. This relies on the spatially varying intensity and, crucially, the polarization texture of structured light: a uniform plane-wave drive can only shift the gap globally, whereas the polarization texture of a skyrmion beam allows the light-induced mass to change sign locally, with an interface geometry set by the beam's spatial structure. Moreover, unlike sample edges or chemically engineered interfaces, these boundaries arise solely from a spatial modulation of the effective Hamiltonian, enabling clean and reconfigurable topological landscapes without modifying the underlying material. Although demonstrated here for a Dirac model, this approach provides a general route toward spatially engineering topological phases with structured light.

\emph{Floquet formalism and effective Hamiltonian}.---To start with, we consider a time-dependent Schr\"odinger equation 
\begin{equation}\label{eq1}
    i\partial_t \ket{\psi(t)} = \mathcal{H}(t) \ket{\psi(t)},
\end{equation}
governed by a time-periodic Hamiltonian $\mathcal{H}(t) = \mathcal{H}(t+T)$, where $T$ denotes the period and $\hbar\omega = 2\pi \hbar/T$ is the corresponding energy scale. By Floquet's theorem \cite{shirley1965,Aoki2016PRB,Rodriguez-Vega_2018,Sota2019Floquet} the solutions of Eq. \eqref{eq1} can be written as $\ket{\psi_\alpha(t)} = e^{-i\varepsilon_\alpha t} \ket{\varphi_\alpha(t)}$, where $\ket{\varphi_\alpha(t)}$ represents the periodic Floquet modes with period $T$, and $\varepsilon_\alpha$ denotes the quasienergies, defined modulo $\omega$.
This formalism allows the time-dependent dynamics to be mapped onto a time-independent eigenvalue problem within an extended Floquet Hilbert space, spanned by $\ket{\mathbbm{u}_\alpha} = \{ \ket{u^{(n)}_\alpha} \}$, where $\ket{u^{(n)}_\alpha}$ (with $n \in \mathbb{Z}$) are the Fourier components of $\ket{\varphi_\alpha(t)}$. This leads to the following eigenvalue problem
\begin{align} \label{eq:matrix-Floquet}
    \mathbbm{H} \ket{\mathbbm{u}_\alpha} = \left( \varepsilon_\alpha + \omega \mathbbm{M} \right) \ket{\mathbbm{u}_\alpha},
\end{align}
where $\mathbbm{H}$ is the Floquet Hamiltonian matrix, with elements $\mathbbm{H}_{n,m}$ given by the $(n-m)$-th Fourier coefficients of $\mathcal{H}(t)$, and 
$\mathbbm{M}_{n,m} = n \delta_{nm}$.

It is usually advantageous to derive an effective Hamiltonian, $\mathbbm{H}_{\rm eff}$, which describes the system's dynamics projected onto a specific subspace, such as the zero-photon sector ($n=0$). This can be accomplished, for example, using Brillouin--Wigner (BW) perturbation theory~\cite{Aoki2016PRB}. By introducing a projector $\mathbbm{P}$ onto the target subspace (e.g., $\mathbbm{P}^{(0)}$ for $n=0$, with $\mathbbm{P}^{(0)}_{n,m} = \delta_{n0}\delta_{m0}$), the effective Hamiltonian then satisfies
\begin{align}
    \mathbbm{H}_{\rm eff} \mathbbm{P} \ket{\mathbbm{u}_\alpha} = \varepsilon_\alpha \mathbbm{P} \ket{\mathbbm{u}_\alpha}.
\end{align}
This effective Hamiltonian encapsulates the essential low-energy dynamics or static properties within the selected subspace, reproducing the quasienergies $\varepsilon_\alpha$ for states predominantly confined to that sector. As detailed in the End Matter, considering a single-frequency time-periodic Hamiltonian ${\cal H}(t)=H_0  + H_1 e^{-i\omega t} + H_{-1} e^{i\omega t}$, 
the effective Hamiltonian reads
$\mathbbm{H}_{\rm eff} = H_0 + \frac{1}{\omega} \left[ H_{-1}, H_1 \right] + \cdots.$

We now consider the Hamiltonian of a two-dimensional Dirac system with mass $m$ and Fermi velocity $v_F$, coupled to a monochromatic electromagnetic field through the vector potential. The time-dependent Hamiltonian is given by
\begin{align}\label{eq:minimal-coupled-Ham}
    \mathcal{H}(t) = v_F \bm{\sigma} \cdot \left[ {\bf p} - \frac{e}{c}\mathbfcal{A}({\bf r},t) \right] + m \sigma_z .
\end{align}
The effective static Hamiltonian $\mathbbm{H}_{\rm eff}$ in the zero-photon sector is obtained using second-order perturbation theory in the coupling $e$, and involves combinations of the Fourier components $H_0$ and $H_{\pm 1}$. Evaluating these contributions and neglecting terms that involve spatial derivatives of the vector potential, i.e.,
$[\bf p, {\bf A}({\bf r})] \approx 0$, we obtain
\begin{align}\label{eq:eff_mass}
    \mathbbm{H}_{\rm eff}
    &\approx 
    v_F^{\rm eff} \bm{\sigma} \cdot {\bf p} + m^{\rm eff} \sigma_z
    \nonumber\\
    &
    +\delta v_1  (\sigma_x p_x-\sigma_y p_y)
    +\delta v_2 (\sigma_x p_y+\sigma_y p_x)
\end{align}
The resulting Hamiltonian consists of a leading term that retains the Dirac form, but with position-dependent renormalizations of both the effective mass and the Fermi velocity:
$m^{\rm eff}=m-4\xi^2\,\big[ m|{\bf A}|^2-\omega  \Im(A_x^\ast A_y)\big]\approx
m + 4\xi^2\,  \Im(A_x^\ast A_y) \,  \omega
$,
$v_F^{\rm eff}/v_F=1-2\xi^2\, |{\bf A}|^2$,
with $\xi=(e/\omega) (v_f/c)$.
The additional terms in the second line of Eq. \eqref{eq:eff_mass}, moreover, introduce anisotropic corrections to the Dirac velocities, with $\delta v_1/v_F=2\xi^2\,\big(|A_x|^2-|A_y|^2\big)$ and $\delta v_2/v_F=4\xi^2\,\Re\big(A_x^\ast A_y\big)$.

\begin{figure}[!]
\centering
\includegraphics[width=0.99\linewidth]{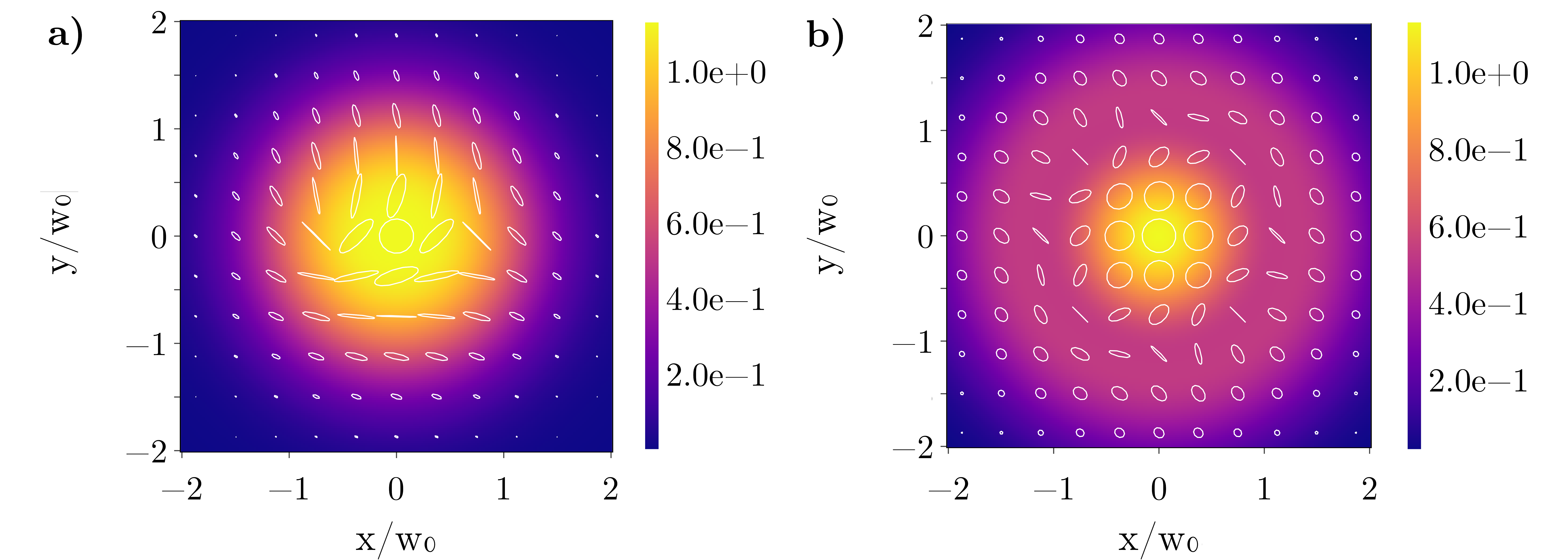}
\caption{Examples of optical skyrmion beams. Color maps show the light intensity, and the overlaid ellipses indicate the local polarization state. The beams are superpositions of two Laguerre--Gaussian (LG) modes with (a) $(\ell_1,\ell_2)=(1,0)$ and (b) $(\ell_1,\ell_2)=(4,0)$, with radial indices $p_1=p_2=0$ in both cases.}
\label{fig2} 
\end{figure}

\emph{Structured light}.---Now, having the general formulation for a spatially modulated Floquet drive, we focus on the impact of OAM-carrying structured light. Specifically, we consider a monochromatic, paraxial beam in the circular polarization basis, as the superposition of Laguerre-Gaussian (LG) beams with different OAM index, i.e.,
\begin{equation}\label{eq7}
    \mathbf{E}(\mathbf{r},t)=\left[LG_0^{\ell_1}(\mathbf{r})\,\hat{\mathbf{L}}+LG_0^{\ell_2}(\mathbf{r})\,\hat{\mathbf{R}}\right]e^{ik(z- ct)},
\end{equation}
where $k=\omega/c$ is the vacuum wave vector , $LG_p^{\ell}(\mathbf{r})$ are LG beams of radial order $p$ and OAM order $\ell$ \cite{oamBook}, and $\{\hat{\mathbf{L}},\hat{\mathbf{R}}\}$ are the left- and right-handed circular polarization basis vectors, respectively \cite{mandelWolf}. Examples of Skyrmion beams of various Skyrmion number are presented in Fig. \ref{fig2}.

The vector potential required for Eq. \eqref{eq:minimal-coupled-Ham} can then be calculated using $\mathbf{E}(\mathbf{r},t)=-i\omega\,\mathbf{A}(\mathbf{r},t)$ \cite{Cohen-Tannoudji1998}. 
Generally, the above electric field describes vector beams, i.e., paraxial beams with nonuniform polarization patterns. For $\Delta\ell=|\ell_1-\ell_2|\neq 0$, however, Eq. \eqref{eq7} describes instead a paraxial Skyrmion, i.e., a light field whose polarization pattern fully covers the Poincar\'e sphere \cite{skyrmionBeams}. Such beams carry a topological invariant, the Skyrmion number, defined as the integer
\begin{equation}\label{eq8}
N=\frac{1}{4\pi}\int\,d^2r\,\left[\mathbf{S}\cdot\left(\frac{\partial\mathbf{S}}{\partial x}\times\frac{\partial\mathbf{S}}{\partial y}\right)\right]_z=\Delta\ell\,F(0,\infty),
\end{equation}
where $\mathbf{S}=\langle\,\mathbf{E}(\mathbf{r},t)|\,\boldsymbol\sigma\,|\mathbf{E}(\mathbf{r},t)\rangle$ is the Stokes vector (with $\boldsymbol\sigma=\{\sigma_x,\sigma_y,\sigma_z\}$ being Pauli matrices), $F(0,\infty)$ accounts for the integral of the radial profile of the beam \cite{skyrmionBeams}, and the integral above is done over the whole Poincar\'e sphere. As it can be seen from Eq. \eqref{eq8}, the primary parameter to tune the Skyrmion number is the difference $\Delta\ell$ in OAM carried by the two LG beams.

\emph{Results}.---As it appears explicitly in Eq. \eqref{eq:eff_mass}, the effective mass term 
depends explicitly on position through the spatial structure of the applied light field and can, in particular, change sign. In the relevant regime $m \ll \omega$, the light-induced mass correction is of order $\delta m \sim (\xi |A|)^2 \,\omega $. For a typical photon energy $\omega\sim 1$ eV and Fermi velocity $5\times 10^5$ m/s, strong yet experimentally accessible electric fields in the range of $0.1-1$ V/nm
yield mass corrections of order $1-100$ meV \cite{DeNinno_2017_Extreme}. Since this energy scale is comparable to typical Dirac masses in TIs, the light-induced correction can become sufficiently large to reverse the sign of the effective mass. This spatially dependent mass renormalization constitutes one of the key results of the present work.

\begin{figure}[!]
\centering
\includegraphics[width=0.99\linewidth]{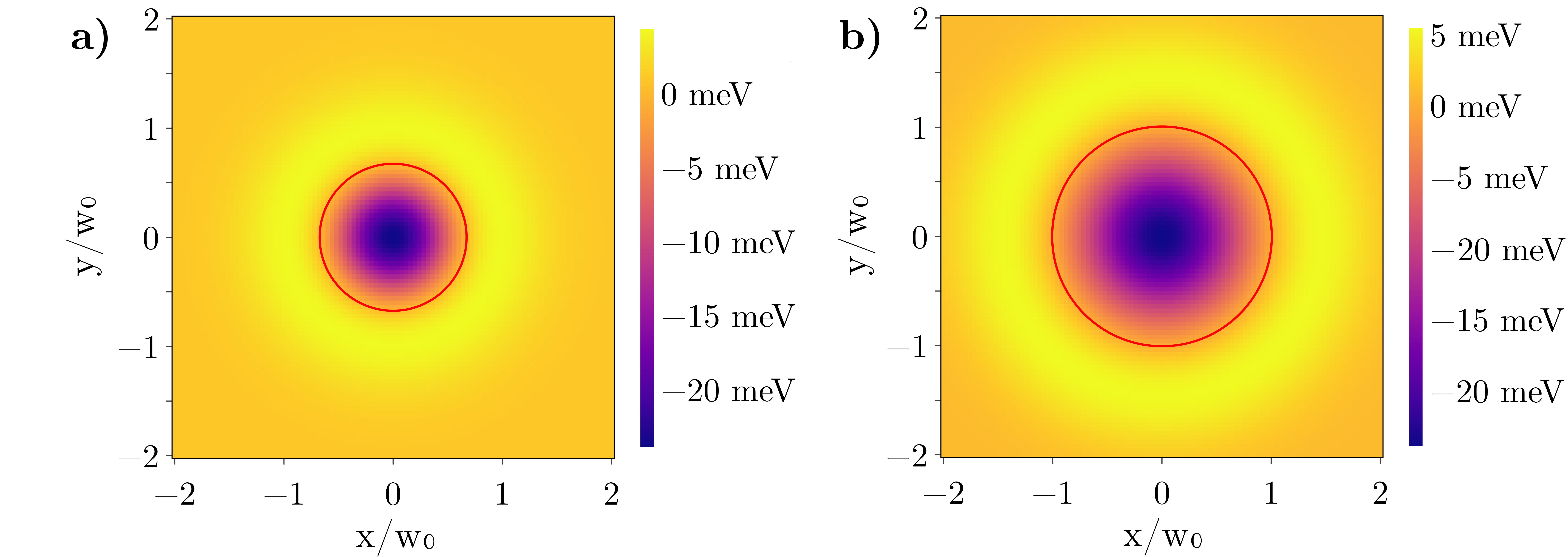}
\caption{ Spatial profile of effective mass $m_{\rm eff}({\bf r})$ induced by the structured light beam for LG modes of $(\ell_1,\ell_2)=(1,0)$ and $(4,0)$, respectively. The contours corresponding to $\delta m=0$ are marked in red. Results are shown for an electric field amplitude \(E_0=3\cdot10^8\) V/m, photon energy \(\hbar \omega = 1\) eV, mass of \(m=1\) meV and a beam waist of \(w_0=1 \; \mu \)m. }
\label{fig3} 
\end{figure}

To illustrate how structured light can spatially control the topological properties of the system, we consider representative Laguerre–Gaussian (LG) modes characterized by different combinations of $(\ell_1,\ell_2)$, whose intensity and polarization textures are shown in Fig.~\ref{fig2}. The corresponding light-induced modulation of the effective Dirac mass is presented in Fig.~\ref{fig3}. Owing to the spatially structured nature of the optical field, the effective mass becomes strongly nonuniform and can change sign across the illuminated region. This produces extended domains with opposite signs of the Dirac mass, separated by contours along which the mass vanishes. Since the sign of the Dirac mass determines the local topological character of the system, these mass-inversion contours define boundaries between topologically distinct regions. They are highlighted by the red curves in Fig.~\ref{fig3}. These results demonstrate that spatially structured light can not only modify the magnitude of the effective mass, but also induce local mass inversions and thereby imprint spatially configurable topological domains and their associated boundaries onto the material. The ability to create light-induced boundaries using structured optical fields has important implications for how interfaces between distinct topological phases can be engineered, as we discuss in more depth later.

Next, we investigate the spin and density profiles induced on a two-dimensional Dirac surface subjected to optical driving by a structured light beam. In the presence of structured light, the effective Hamiltonian contains spatially varying mass and velocity terms, making a numerical treatment necessary to determine the resulting spin and density distributions. Since the characteristic length scale of the spatial variations induced by the structured light is much larger than the electronic wavelength and thus the effective Hamiltonian varies slowly in real space, we employ a \emph{Wigner transformation}, which provides a systematic connection between the locally homogeneous approximation and corrections arising from spatial gradients (see End Matter for more details). After obtaining the leading-order Green's function and its first-order gradient correction, the relevant physical observables can be evaluated by integrating over momentum space. In particular, the local spin-resolved spectral density at position $\mathbf{R}$ and energy $\epsilon$ is given by
\begin{equation}\label{eq:spin_density}
\langle \sigma^a \rangle(\mathbf{R};\epsilon) = -\frac{1}{\pi} \Im \int \frac{d^2k}{(2\pi)^2} \, \text{Tr} \left[ \sigma^a G_W(\mathbf{R}, \mathbf{k}; \epsilon) \right],
\end{equation}
where $G_W$ denotes the Wigner-transformed Green's function, including the gradient corrections \cite{Patel2023,Megias2021}. This formulation captures the local spin structure generated by the spatial modulation of the effective Hamiltonian, in particular by variations of the light-induced mass and velocity. The corresponding density of states and spin polarization can then be obtained from these local spectral quantities and, when appropriate, integrated over energy with the relevant occupation factor.

\begin{figure}[t]
\centering
\includegraphics[width=0.99\linewidth]{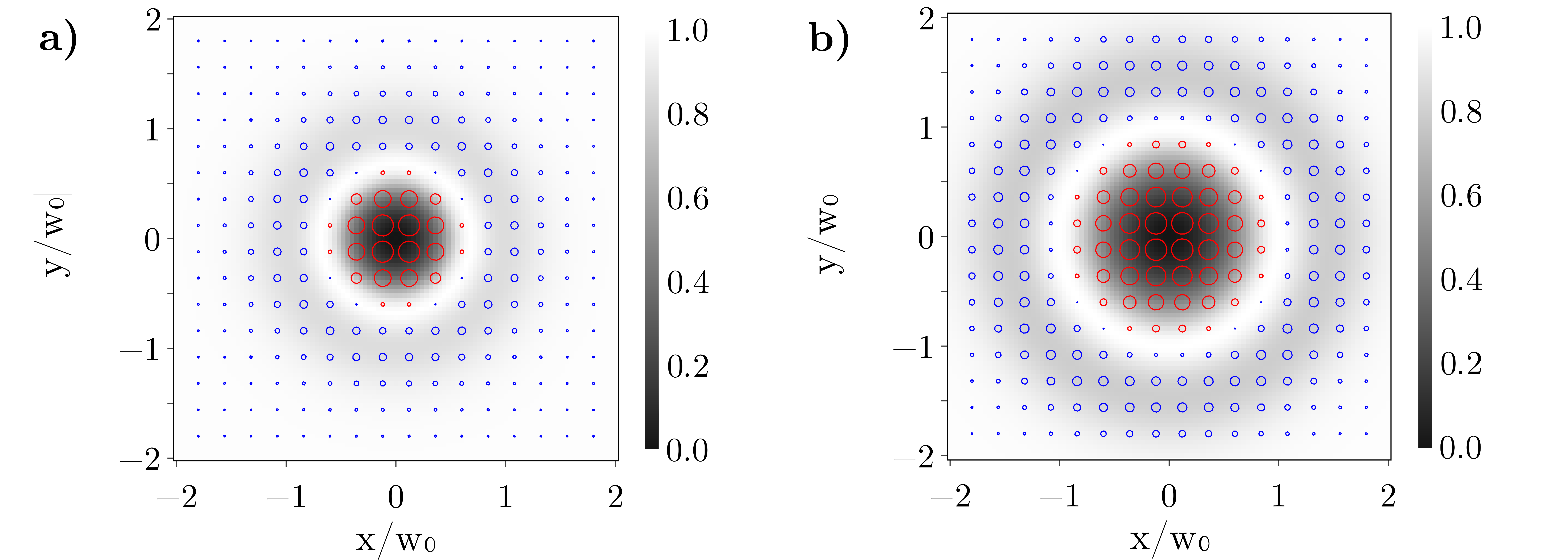}
\caption{The resulting probability density and spin structure induced by the structured light beam for LG modes of 
(a) $(\ell_1,\ell_2)=(1,0)$ and (b) $(4,0)$. The blue colored ellipses correspond to negative $z$-spin and red to positive $z$-spin while the shape represents a tilted circle in the direction of the $x$- and $y$-spins.
}
\label{fig4} 
\end{figure}

The resulting probability density and spin texture induced by the structured light corresponding to the vector beams in Fig. \ref{fig2} are shown in Fig. \ref{fig4}. The density profile largely follows the spatial intensity profile of the light, while the induced spin-resolved densities closely track the spatially varying effective mass generated by the light--matter interaction. In particular, the down-spin (up-spin) density is enhanced in regions where the effective mass is positive (negative). This behavior can be understood from the local spin texture of the Floquet Hamiltonian, whose spin polarization is controlled by the sign of the effective mass. Since the optical field varies smoothly on the relevant electronic length scales, the system is locally well approximated by a spatially uniform Hamiltonian with parameters set by the local optical field.

\emph{Discussion}.---We first discuss why light-induced boundaries between topologically distinct regions are appealing from both a conceptual and an experimental perspective. In conventional solid-state realizations, topological boundaries are defined either by the physical termination of a sample or by interfaces between materials with different compositions or band structures. Such boundaries inevitably inherit structural disorder, interface roughness, chemical inhomogeneity, and other fabrication-related imperfections, which can degrade boundary-state transport and cause deviations from the expected quantized response \cite{Knez2014}. 
In contrast, optically induced boundaries are defined by the spatial profile of the driving field and are therefore free from random roughness and local distortions, while remaining smooth, continuously tunable, and dynamically reconfigurable. Combined with the established ability of Floquet engineering to modify band topology \cite{lindner2011floquet,Cayssol2013,rudner2020band}, spatially structured light may thus provide a flexible platform for creating and manipulating topological interfaces. We note that, while spatially modulated driving has been shown to bind localized modes in Floquet topological insulators \cite{Katan2013PRL,Katan2013PRB}, the use of topologically structured light to design and reconfigure topological landscapes in matter has remained largely unexplored.

We further find that the local spin density obtained from the Wigner-gradient expansion closely follows the spatial profile of the light-induced mass. This is consistent with an adiabatic picture: since the optical modulation varies slowly compared with the relevant electronic length scales, the electronic states locally follow the spatially varying effective Hamiltonian. The resulting spin texture suggests that more complex optical textures could be imprinted onto electronic degrees of freedom. In our simplified model, however, the induced spin texture carries no net skyrmion number and therefore does not constitute an electronic skyrmion. More general models supporting skyrmionic states, together with treatments beyond the perturbative regime, may provide a route toward transferring topological textures between light and electronic systems, in line with a related recent study \cite{Asmar2025}.

Finally, beyond the spatially modulated topological mass and the resulting boundary and spin textures, the light--matter interaction also renormalizes the velocity terms and generates additional anisotropic contributions. These provide further possibilities for tailoring the driven topological phases. In particular, spatially varying velocity terms can be interpreted in terms of position-dependent local frame fields and an effective geometry, suggesting a possible route toward engineering emergent curved-spacetime geometries for electronic quasiparticles using structured light.

\emph{Conclusions}.---In this work, we have shown that spatially structured light provides a versatile route to locally engineer topological phases of matter. In particular, optically defined topological boundaries offer a reconfigurable alternative to sample terminations and chemically engineered interfaces, while avoiding the intrinsic structural and compositional imperfections associated with such material boundaries. By imprinting its spatial structure onto the effective Hamiltonian, structured light can create and reshape topological boundaries, induce characteristic local spin textures, and provide additional control through velocity and anisotropy renormalization. These results highlight spatial Floquet engineering as a promising framework for dynamically reconfigurable topological systems and motivate further exploration of the interplay and possible transfer of topological structures between light and matter.

\emph{Acknowledgements}.---M.M. acknowledges financial support from the Magnus Ehrnrooth Foundation and the Finnish Cultural Foundation, Pirkanmaa Regional Fund. 
A.G.M. acknowledges the financial support from the Finnish Center of Excellence in Quantum Materials QMAT (No. 374166), and the ERC Consolidator Grant ULTRATWISTROICS (No. 101170477). T.O. acknowledges support from the Finnish Research Council project 362573 and the Finnish quantum flagship program.  M.O. acknowledges the financial support from the Research Council of Finland Photonics Research and Innovation Flagship (PREIN: decision No. 320165).
This work is part of
the Finnish Center of Excellence in Quantum Materials (QMAT).

\bibliography{refs}

@article{Aoki2016PRB,
  title = {Brillouin-Wigner theory for high-frequency expansion in periodically driven systems: Application to Floquet topological insulators},
  author = {Mikami, Takahiro and Kitamura, Sota and Yasuda, Kenji and Tsuji, Naoto and Oka, Takashi and Aoki, Hideo},
  journal = {Phys. Rev. B},
  volume = {93},
  issue = {14},
  pages = {144307},
  numpages = {25},
  year = {2016},
  month = {Apr},
  publisher = {American Physical Society},
  doi = {10.1103/PhysRevB.93.144307},
  url = {https://link.aps.org/doi/10.1103/PhysRevB.93.144307}
}

@article{Katan2013PRB,
  title = {Generation and manipulation of localized modes in Floquet topological insulators},
  author = {Tenenbaum Katan, Yaniv and Podolsky, Daniel},
  journal = {Phys. Rev. B},
  volume = {88},
  issue = {22},
  pages = {224106},
  numpages = {10},
  year = {2013},
  month = {Dec},
  publisher = {American Physical Society},
  doi = {10.1103/PhysRevB.88.224106},
  url = {https://link.aps.org/doi/10.1103/PhysRevB.88.224106}
}

@article{Asmar2025,
  title = {Photoinduced multiply quantized vortex states in Dirac-like materials},
  author = {Massaro, Lauren I. and Meese, Connor and Sandler, Nancy P. and Asmar, Mahmoud M.},
  journal = {Phys. Rev. B},
  volume = {111},
  issue = {8},
  pages = {085402},
  numpages = {13},
  year = {2025},
  month = {Feb},
  publisher = {American Physical Society},
  doi = {10.1103/PhysRevB.111.085402},
  url = {https://link.aps.org/doi/10.1103/PhysRevB.111.085402}
}

@article{Katan2013PRL,
  title = {Modulated Floquet Topological Insulators},
  author = {Katan, Yaniv Tenenbaum and Podolsky, Daniel},
  journal = {Phys. Rev. Lett.},
  volume = {110},
  issue = {1},
  pages = {016802},
  numpages = {5},
  year = {2013},
  month = {Jan},
  publisher = {American Physical Society},
  doi = {10.1103/PhysRevLett.110.016802},
  url = {https://link.aps.org/doi/10.1103/PhysRevLett.110.016802}
}

@article{Kitagawa2010,
  title = {Topological characterization of periodically driven quantum systems},
  author = {Kitagawa, Takuya and Berg, Erez and Rudner, Mark and Demler, Eugene},
  journal = {Phys. Rev. B},
  volume = {82},
  issue = {23},
  pages = {235114},
  numpages = {12},
  year = {2010},
  month = {Dec},
  publisher = {American Physical Society},
  doi = {10.1103/PhysRevB.82.235114},
  url = {https://link.aps.org/doi/10.1103/PhysRevB.82.235114}
}

@article{Rudner_2015,
doi = {10.1088/1367-2630/17/12/125014},
url = {https://doi.org/10.1088/1367-2630/17/12/125014},
year = {2015},
month = {dec},
publisher = {IOP Publishing},
volume = {17},
number = {12},
pages = {125014},
author = {Nathan, Frederik and Rudner, Mark S},
title = {Topological singularities and the general classification of Floquet–Bloch systems},
journal = {New Journal of Physics}
}

@article{Cayssol2013,
author = {Cayssol, Jérôme and Dóra, Balázs and Simon, Ferenc and Moessner, Roderich},
title = {Floquet topological insulators},
journal = {physica status solidi (RRL) – Rapid Research Letters},
volume = {7},
number = {1-2},
pages = {101-108},
doi = {https://doi.org/10.1002/pssr.201206451},
year = {2013}
}

@article{rudner2020band,
  title={Band structure engineering and non-equilibrium dynamics in Floquet topological insulators},
  author={Rudner, Mark S and Lindner, Netanel H},
  journal={Nature reviews physics},
  volume={2},
  number={5},
  pages={229--244},
  year={2020},
  doi={https://doi.org/10.1038/s42254-020-0170-z},
  publisher={Nature Publishing Group UK London}
}

@article{Knez2014,
  title = {Observation of Edge Transport in the Disordered Regime of Topologically Insulating $\mathrm{InAs}/\mathrm{GaSb}$ Quantum Wells},
  author = {Knez, Ivan and Rettner, Charles T. and Yang, See-Hun and Parkin, Stuart S. P. and Du, Lingjie and Du, Rui-Rui and Sullivan, Gerard},
  journal = {Phys. Rev. Lett.},
  volume = {112},
  issue = {2},
  pages = {026602},
  numpages = {5},
  year = {2014},
  month = {Jan},
  publisher = {American Physical Society},
  doi = {10.1103/PhysRevLett.112.026602},
  url = {https://link.aps.org/doi/10.1103/PhysRevLett.112.026602}
}

@book{Cohen-Tannoudji1998,
  title={Atom-photon interactions: basic processes and applications},
  author={Cohen-Tannoudji, Claude and Dupont-Roc, Jacques and Grynberg, Gilbert},
  year={1998},
  publisher={John Wiley \& Sons}
}

@article{rubio2018lightmatter,
  title={From a quantum-electrodynamical light--matter description to novel spectroscopies},
  author={Ruggenthaler, Michael and Tancogne-Dejean, Nicolas and Flick, Johannes and Appel, Heiko and Rubio, Angel},
  journal={Nat. Rev. Chem.},
  volume={2},
  number={3},
  pages={1--16},
  doi = {10.1038/s41570-018-0118},
  year={2018},
  publisher={Nature Publishing Group}
}

@article{nori2019lightmatter,
  title={Ultrastrong coupling between light and matter},
  author={Frisk Kockum, Anton and Miranowicz, Adam and De Liberato, Simone and Savasta, Salvatore and Nori, Franco},
  journal={Nature Reviews Physics},
  volume={1},
  number={1},
  pages={19--40},
  year={2019},
  doi = {10.1038/s42254-018-0006-2},
  publisher={Nature Publishing Group UK London}
}

@article{rivera2020lightmatter,
  title={Light--matter interactions with photonic quasiparticles},
  author={Rivera, Nicholas and Kaminer, Ido},
  journal={Nat. Rev. Phys.},
  volume={2},
  number={10},
  pages={538--561},
  year={2020},
  doi = {10.1038/s42254-020-0224-2},
  publisher={Nature Publishing Group UK London}
}

@article{gutzler2021lightmatter,
  title={Light--matter interaction at atomic scales},
  author={Gutzler, Rico and Garg, Manish and Ast, Christian R and Kuhnke, Klaus and Kern, Klaus},
  journal={Nat. Rev. Phys.},
  volume={3},
  number={6},
  pages={441},
  doi = {10.1038/s42254-021-00306-5},
  year={2021},
  publisher={Nature Publishing Group UK London}
}

@article{Ultrastrong2019lightmatter,
  title = {Ultrastrong coupling regimes of light-matter interaction},
  author = {Forn-D\'{\i}az, P. and Lamata, L. and Rico, E. and Kono, J. and Solano, E.},
  journal = {Rev. Mod. Phys.},
  volume = {91},
  issue = {2},
  pages = {025005},
  numpages = {48},
  year = {2019},
  month = {Jun},
  publisher = {American Physical Society},
  doi = {10.1103/RevModPhys.91.025005},
  url = {https://link.aps.org/doi/10.1103/RevModPhys.91.025005}
}

@article{bloch2022strongly,
  title={Strongly correlated electron--photon systems},
  author={Bloch, Jacqueline and Cavalleri, Andrea and Galitski, Victor and Hafezi, Mohammad and Rubio, Angel},
  journal={Nature},
  volume={606},
  number={7912},
  pages={41--48},
  year={2022},
  doi = {10.1038/s41586-022-04726-w},
  publisher={Nature Publishing Group UK London}
}

@article{mitrano2016superconductivity,
  title={Possible light-induced superconductivity in K3C60 at high temperature},
  author={Mitrano, Matteo and Cantaluppi, Alice and Nicoletti, Daniele and Kaiser, Stefan and Perucchi, A and Lupi, Stefano and Di Pietro, P and Pontiroli, Daniele and Ricc{\`o}, Mauro and Clark, Stephen R and others},
  journal={Nature},
  volume={530},
  number={7591},
  pages={461--464},
  year={2016},
  doi = {10.1038/nature16522},
  publisher={Nature Publishing Group UK London}
}

@article{demsar2016light,
  title={Light-induced superconductivity},
  author={Demsar, Jure},
  journal={Nat. Phys.},
  volume={12},
  number={3},
  pages={202--203},
  year={2016},
  doi = {10.1038/nphys3687},
  publisher={Nature Publishing Group UK London}
}

@article{Aoki2009floquet,
  title = {Photovoltaic Hall effect in graphene},
  author = {Oka, Takashi and Aoki, Hideo},
  journal = {Phys. Rev. B},
  volume = {79},
  issue = {8},
  pages = {081406},
  numpages = {4},
  year = {2009},
  month = {Feb},
  publisher = {American Physical Society},
  doi = {10.1103/PhysRevB.79.081406},
  url = {https://link.aps.org/doi/10.1103/PhysRevB.79.081406}
}

@article{Demler2011floquet,
  title = {Transport properties of nonequilibrium systems under the application of light: Photoinduced quantum Hall insulators without Landau levels},
  author = {Kitagawa, Takuya and Oka, Takashi and Brataas, Arne and Fu, Liang and Demler, Eugene},
  journal = {Phys. Rev. B},
  volume = {84},
  issue = {23},
  pages = {235108},
  numpages = {13},
  year = {2011},
  month = {Dec},
  publisher = {American Physical Society},
  doi = {10.1103/PhysRevB.84.235108},
  url = {https://link.aps.org/doi/10.1103/PhysRevB.84.235108}
}

@article{lindner2011floquet,
  title={Floquet topological insulator in semiconductor quantum wells},
  author={Lindner, Netanel H and Refael, Gil and Galitski, Victor},
  journal={Nat. Phys.},
  volume={7},
  number={6},
  pages={490--495},
  year={2011},
  doi = {10.1038/nphys1926},
  publisher={Nature Publishing Group UK London}
}

@article{rechtsman2013floquet,
  title={Photonic Floquet topological insulators},
  author={Rechtsman, Mikael C and Zeuner, Julia M and Plotnik, Yonatan and Lumer, Yaakov and Podolsky, Daniel and Dreisow, Felix and Nolte, Stefan and Segev, Mordechai and Szameit, Alexander},
  journal={Nature},
  volume={496},
  number={7444},
  pages={196--200},
  year={2013},
  doi = {10.1038/nature12066},
  publisher={Nature Publishing Group UK London}
}

@article{Dehghani2015floquet,
  title = {Out-of-equilibrium electrons and the Hall conductance of a Floquet topological insulator},
  author = {Dehghani, Hossein and Oka, Takashi and Mitra, Aditi},
  journal = {Phys. Rev. B},
  volume = {91},
  issue = {15},
  pages = {155422},
  numpages = {11},
  year = {2015},
  month = {Apr},
  publisher = {American Physical Society},
  doi = {10.1103/PhysRevB.91.155422},
  url = {https://link.aps.org/doi/10.1103/PhysRevB.91.155422}
}

@article{mciver2020floquet,
  title={Light-induced anomalous Hall effect in graphene},
  author={McIver, James W and Schulte, Benedikt and Stein, F-U and Matsuyama, Toru and Jotzu, Gregor and Meier, Guido and Cavalleri, Andrea},
  journal={Nat. Phys.},
  volume={16},
  number={1},
  pages={38--41},
  year={2020},
  doi = {10.1038/s41567-019-0698-y},
  publisher={Nature Publishing Group UK London}
}

@inproceedings{sentef2024floquet,
  title={Light-matter control of quantum materials: from Floquet to cavity engineering},
  author={Sentef, Michael A},
  booktitle={Physical Chemistry of Semiconductor Materials and Interfaces XXIII},
  pages={PC131270P},
  year={2024},
  organization={SPIE}
}

@article{merboldt2024floquet,
  title={Observation of Floquet states in graphene},
  author={Merboldt, Marco and Sch{\"u}ler, Michael and Schmitt, David and Bange, Jan Philipp and Bennecke, Wiebke and Gadge, Karun and Pierz, Klaus and Schumacher, Hans Werner and Momeni, Davood and Steil, Daniel and others},
  journal={arXiv:2404.12791},
  url={https://arxiv.org/abs/2404.12791},
  year={2024}
}

@article{ZHANG2020135197,
title = {Feynman rules in terms of the Wigner transformed Green functions},
journal = {Physics Letters B},
volume = {802},
pages = {135197},
year = {2020},
issn = {0370-2693},
doi = {https://doi.org/10.1016/j.physletb.2020.135197},
url = {https://www.sciencedirect.com/science/article/pii/S0370269320300010},
author = {C.X. Zhang and M.A. Zubkov}
}

@article{Von-Eiff1992,
  title = {Approach to the relativistic extended Thomas-Fermi expansion for Green's functions, phase-space densities, and densities},
  author = {Von-Eiff, D. and Haddad, S. and Weigel, M. K.},
  journal = {Phys. Rev. C},
  volume = {46},
  issue = {1},
  pages = {230--237},
  numpages = {0},
  year = {1992},
  month = {Jul},
  publisher = {American Physical Society},
  doi = {10.1103/PhysRevC.46.230},
  url = {https://link.aps.org/doi/10.1103/PhysRevC.46.230}
}

@article{Megias2021,
  title = {Analytical Green’s functions for continuum spectra},
  author = {Megías, Eugenio and Quirós, Mariano},
  journal = {Journal of High Energy Physics},
  volume = {2021},
  issue = {9},
  year = {2021},
  doi = {10.1007/JHEP09(2021)157},
  url = {https://doi.org/10.1007/JHEP09(2021)157}
}

@article{Patel2023,
doi = {10.1007/s41745-023-00406-4},
url = {https://doi.org/10.1007/s41745-023-00406-4},
year = {2023},
month = {apr},
volume = {103},
issue = {2},
pages = {401--417},
author = {Patel, Apoorva D.},
title = {The Quantum Density Matrix and Its Many Uses},
journal = {Journal of the Indian Institute of Science}
}

@book{oamBook,
  title={The Angular Momentum of Light},
  editor={Andrews, David and Babiker, Mohammed},
  year={2013},
  publisher={Cambridge}
}

@book{mandelWolf,
  title={Optical Coherence and Quantum Optics},
  author={Mandel, Leonard and Wolf, Emil},
  year={1995},
  publisher={Cambridge}
}

@article{skyrmionBeams,
  title = {Paraxial skyrmionic beams},
  author = {Gao, Sijia and Speirits, Fiona C. and Castellucci, Francesco and Franke-Arnold, Sonja and Barnett, Stephen M. and G\"otte, J\"org B.},
  journal = {Phys. Rev. A},
  volume = {102},
  issue = {5},
  pages = {053513},
  numpages = {6},
  year = {2020},
  month = {Nov},
  publisher = {American Physical Society},
  doi = {10.1103/PhysRevA.102.053513},
  url = {https://link.aps.org/doi/10.1103/PhysRevA.102.053513}
}

@book{ref1,
	title = {The Angular Momentum of Light},
	publisher = {Cambridge University Press},
	editor = {Andrews, D. L. and Babiker, M.},
	year = {2013},
}

@incollection{Dennis_Singular_optics_2009,
title = {Singular Optics: Optical Vortices and Polarization Singularities},
editor = {E Wolf},
series = {Progress in Optics},
publisher = {Elsevier},
volume = {53},
pages = {293-363},
year = {2009},
issn = {0079-6638},
doi = {https://doi.org/10.1016/S0079-6638(08)00205-9},
author = {Mark R. Dennis and Kevin O'Holleran and Miles J. Padgett}
}

@article{ref4,
  title={Optical skyrmions and other topological quasiparticles of light},
  author={Shen, Y. and Zhang, Q. and Shi, P. and Du, L. and Yuan, X. and Zayats, A. V.},
  journal={Nat. Photon.},
  volume={18},
  pages={15},
  year={2024},
  url={https://doi.org/10.1038/s41566-023-01325-7}
}

@article{Rosales-Guzman_2018,
doi = {10.1088/2040-8986/aaeb7d},
year = {2018},
month = {nov},
publisher = {IOP Publishing},
volume = {20},
number = {12},
pages = {123001},
author = {Rosales-Guzmán, Carmelo and Ndagano, Bienvenu and Forbes, Andrew},
title = {A review of complex vector light fields and their applications},
journal = {J. Opt.},
}

@book{forbesBook,
	title = {Laser beam propagation - generation and propagation of customised light},
	publisher = {CRC Press},
	editor = {Forbes, A.},
	year = {2014},
}

@article{OAMvsTI,
  title = {Spin and charge transport induced by a twisted light beam on the surface of a topological insulator},
  author = {Shintani, Kunitaka and Taguchi, Katsuhisa and Tanaka, Yukio and Kawaguchi, Yuki},
  journal = {Phys. Rev. B},
  volume = {93},
  issue = {19},
  pages = {195415},
  numpages = {22},
  year = {2016},
  month = {May},
  publisher = {American Physical Society},
  doi = {10.1103/PhysRevB.93.195415},
  url = {https://link.aps.org/doi/10.1103/PhysRevB.93.195415}
}

@article{ChernCircular,
  title = {Fermionic Chern insulator from twisted light with linear polarization},
  author = {Bhattacharya, Utso and Chaudhary, Swati and Grass, Tobias and Johnson, Allan S. and Wall, Simon and Lewenstein, Maciej},
  journal = {Phys. Rev. B},
  volume = {105},
  issue = {8},
  pages = {L081406},
  numpages = {6},
  year = {2022},
  month = {Feb},
  publisher = {American Physical Society},
  doi = {10.1103/PhysRevB.105.L081406},
  url = {https://link.aps.org/doi/10.1103/PhysRevB.105.L081406}
}

@article{frisk_kockum_ultrastrong_2019,
  title   = {Ultrastrong coupling regimes of light-matter interaction},
  author  = {Frisk Kockum, A. and Miranowicz, A. and Liberato, S. and Savasta, S. and Nori, F.},
  journal = {Reviews of Modern Physics},
  year    = {2019},
  volume  = {91},
  number  = {3},
  pages   = {035005},
  doi     = {10.1103/RevModPhys.91.035005}
}

@article{hubener_quantum_2024,
  title   = {Quantum materials engineering by structured cavity vacuum fluctuations},
  author  = {H{\"u}bener, H. and Bostr{\"o}m, E. V. and Claassen, M. and Latini, S. and Rubio, A.},
  journal = {Materials for Quantum Technology},
  year    = {2024},
  volume  = {4},
  number  = {2},
  pages   = {023002},
  doi     = {10.1088/2633-4356/ad3c3e}
}

@article{Stammer_PRXQuantum,
  title = {Quantum Electrodynamics of Intense Laser-Matter Interactions: A Tool for Quantum State Engineering},
  author = {Stammer, Philipp and Rivera-Dean, Javier and Maxwell, Andrew and Lamprou, Theocharis and Ord\'o\~nez, Andr\'es and Ciappina, Marcelo F. and Tzallas, Paraskevas and Lewenstein, Maciej},
  journal = {PRX Quantum},
  volume = {4},
  issue = {1},
  pages = {010201},
  numpages = {43},
  year = {2023},
  month = {Jan},
  publisher = {American Physical Society},
  doi = {10.1103/PRXQuantum.4.010201},
  url = {https://link.aps.org/doi/10.1103/PRXQuantum.4.010201}
}

@article{svendsen_combining_2021,
  title   = {Combining density functional theory with macroscopic {QED} for quantum light--matter interactions in 2{D} materials},
  author  = {Svendsen, M. K. and {et al.}},
  journal = {Nature Communications},
  year    = {2021},
  volume  = {12},
  pages   = {2778},
  doi     = {10.1038/s41467-021-23062-0}
}

@article{shirley1965,
  title={Solution of the Schr{\"o}dinger equation with a Hamiltonian periodic in time},
  author={Shirley, Jon H},
  journal={Physical Review},
  volume={138},
  number={4B},
  pages={B979},
  year={1965},
  doi={https://doi.org/10.1103/PhysRev.138.B979},
  publisher={APS}
}

@article{Sota2019Floquet,
   author = "Oka, Takashi and Kitamura, Sota",
   title = "Floquet Engineering of Quantum Materials", 
   journal= "Annual Review of Condensed Matter Physics",
   year = "2019",
   volume = "10",
   number = "Volume 10, 2019",
   pages = "387-408",
   doi = "https://doi.org/10.1146/annurev-conmatphys-031218-013423"
  }

@article{Rodriguez-Vega_2018,
doi = {10.1088/1367-2630/aade37},
url = {https://doi.org/10.1088/1367-2630/aade37},
year = {2018},
month = {sep},
publisher = {IOP Publishing},
volume = {20},
number = {9},
pages = {093022},
author = {Rodriguez-Vega, M and Lentz, M and Seradjeh, B},
title = {Floquet perturbation theory: formalism and application to low-frequency limit},
journal = {New Journal of Physics}
}

@article{stoyanov2023optical,
  title={Optical vortices in brief: introduction for experimentalists},
  author={Stoyanov, Lyubomir and Topuzoski, Suzana and Paulus, Gerhard G and Dreischuh, Alexander},
  journal={The European Physical Journal Plus},
  volume={138},
  number={8},
  pages={702},
  year={2023},
  publisher={Springer}
}

@article{he2024optical,
  title={Optical skyrmions from metafibers with subwavelength features},
  author={He, Tiantian and Meng, Yuan and Wang, Lele and Zhong, Hongkun and Mata-Cervera, Nilo and Li, Dan and Yan, Ping and Liu, Qiang and Shen, Yijie and Xiao, Qirong},
  journal={Nature Communications},
  volume={15},
  number={1},
  pages={10141},
  year={2024},
  doi={https://doi.org/10.1038/s41467-024-54207-z},
  publisher={Nature Publishing Group UK London}
}

@article{du2019deep,
  title={Deep-subwavelength features of photonic skyrmions in a confined electromagnetic field with orbital angular momentum},
  author={Du, Luping and Yang, Aiping and Zayats, Anatoly V and Yuan, Xiaocong},
  journal={Nature Physics},
  volume={15},
  number={7},
  pages={650--654},
  year={2019},
  publisher={Nature Publishing Group UK London}
}

@article{DeNinno_2017_Extreme,
  title = {Extreme-Ultraviolet Vortices from a Free-Electron Laser},
  author = {Rebernik Ribi\ifmmode \check{c}\else \v{c}\fi{}, Primo\ifmmode \check{z}\else \v{z}\fi{} and R\"osner, Benedikt and Gauthier, David and Allaria, Enrico and D\"oring, Florian and Foglia, Laura and Giannessi, Luca and Mahne, Nicola and Manfredda, Michele and Masciovecchio, Claudio and Mincigrucci, Riccardo and Mirian, Najmeh and Principi, Emiliano and Roussel, El\'eonore and Simoncig, Alberto and Spampinati, Simone and David, Christian and De Ninno, Giovanni},
  journal = {Phys. Rev. X},
  volume = {7},
  issue = {3},
  pages = {031036},
  numpages = {9},
  year = {2017},
  month = {Aug},
  publisher = {American Physical Society},
  doi = {10.1103/PhysRevX.7.031036},
  url = {https://link.aps.org/doi/10.1103/PhysRevX.7.031036}
}

@article{lininger2023chirality,
  title={Chirality in light--matter interaction},
  author={Lininger, Andrew and Palermo, Giovanna and Guglielmelli, Alexa and Nicoletta, Giuseppe and Goel, Madhav and Hinczewski, Michael and Strangi, Giuseppe},
  journal={Advanced Materials},
  volume={35},
  number={34},
  pages={2107325},
  year={2023},
  publisher={Wiley Online Library}
}

@article{appugliese_breakdown_2022,
  title   = {Breakdown of topological protection by cavity vacuum fields in the integer quantum {H}all effect},
  author  = {Appugliese, F. and {et al.}},
  journal = {Science},
  year    = {2022},
  volume  = {375},
  number  = {6584},
  pages   = {1030--1034},
  doi     = {10.1126/science.abl6188}
}

@article{hubener_engineering_2021,
  title   = {Engineering quantum materials with chiral optical cavities},
  author  = {H{\"u}bener, H. and {et al.}},
  journal = {Nature Materials},
  year    = {2021},
  volume  = {20},
  pages   = {438--442},
  doi     = {10.1038/s41563-020-00906-9}
}

@article{goto_chern_2025,
  title   = {Chern insulators and topological flat bands in cavity-embedded kagome systems},
  author  = {Goto, H. and Okugawa, R. and Tohyama, T.},
  journal = {arXiv preprint},
  year    = {2025},
  eprint  = {2510.09150},
  archivePrefix = {arXiv},
  primaryClass  = {cond-mat.mes-hall}
}

@article{oka_floquet_2019,
  title   = {Floquet engineering of quantum materials},
  author  = {Oka, T. and Kitamura, S.},
  journal = {Annual Review of Condensed Matter Physics},
  year    = {2019},
  volume  = {10},
  pages   = {387--408},
  doi     = {10.1146/annurev-conmatphys-031218-013423}
}

@article{floquet_topo_exp_2026,
  title   = {First direct experimental evidence of a {F}loquet topological state},
  author  = {{et al.}},
  journal = {Nature Physics},
  year    = {2026},
  note    = {Experimental observation of a light-induced Floquet topological insulator}
}

\section{End Matter}

We start from the time-dependent Schr\"odinger equation 
$i\partial_t \ket{\psi(t)} = {\cal H}(t) \ket{\psi(t)}$
for a single particle
with a time-periodic Hamiltonian ${\cal H}(t)={\cal H}(t+T)$.
According to Floquet's theorem, the solutions can be written as
\begin{align}
    \ket{\psi_\alpha(t)} = e^{-i\varepsilon_\alpha t} \ket{\varphi_\alpha(t)}
    = e^{-i\varepsilon_\alpha t} \sum_{n} e^{-i n\omega t} \ket{u^{(n)}_\alpha},
\end{align}
where $\varepsilon_\alpha$ are the quasienergies,
$\omega = 2\pi/T$ is the driving frequency, and
$\ket{u^{(n)}_\alpha}$ are the Floquet eigenstates.
As in Bloch theory, the quasienergies can be restricted to the first Floquet
Brillouin zone, e.g., $0 \leq \varepsilon_\alpha < 2\pi/T$.
The Floquet eigenstates satisfy
\begin{align}
    \left(\varepsilon_\alpha + n\omega \right) \ket{u^{(n)}_\alpha}
    = \sum_{m} \mathbbm{H}_{n,m} \ket{u^{(m)}_\alpha},
\end{align}
with the Fourier components of the Hamiltonian given by
$ \mathbbm{H}_{n,m}
    = (1/T)\int_0^T dt \: e^{i(n-m)\omega t}\: {\cal H}(t)$.  
Introducing $\ket{\mathbbm{u}_\alpha}=\{\ket{u^{(n)}_\alpha}\}$, the eigenvalue
problem can be recast in matrix form as
\begin{align} \label{eq:matrix-Floquet}
    \left( \mathbbm{H}- \varepsilon_\alpha \right)
    \ket{\mathbbm{u}_\alpha}
    = \omega\, \mathbbm{M} \ket{\mathbbm{u}_\alpha},
\end{align}
where $\mathbbm{M}_{n,m} = n\delta_{nm}$.

Following Ref.~\cite{Aoki2016PRB}, we employ BW theory to construct
an effective Hamiltonian by projecting onto a chosen subspace:
\begin{align}
    \mathbbm{H}_{\rm eff}\,\mathbbm{P}\ket{\mathbbm{u}_\alpha}
    = \varepsilon_\alpha\, \mathbbm{P}\ket{\mathbbm{u}_\alpha},
\end{align}
where $\mathbbm{P}$ is the projector.
We focus on the static effective Hamiltonian obtained by projecting onto the
zero-photon sector, with
$\mathbbm{P}^{(0)}_{n,m} = \delta_{n0}\delta_{m0}$,
This projector commutes with $\mathbbm{M}$ and satisfies
$\mathbbm{P}^{(0)}\mathbbm{M}=0$.
Projecting Eq.~\eqref{eq:matrix-Floquet} onto the 
complementary subspace with 
$\mathbbm{Q}=\mathbbm{1}-\mathbbm{P}^{(0)}$ and using the
invertibility of $\omega\mathbbm{M}\mathbbm{Q}$, we obtain
\begin{align} \label{eq:pre-new-resolvent}
    \mathbbm{Q} \ket{\mathbbm{u}_\alpha}
    = \frac{\mathbbm{Q}}{\omega \mathbbm{M}}
    (\mathbbm{H}-\varepsilon)\ket{\mathbbm{u}_\alpha}.
\end{align}
We can thus define the resolvent operator
\begin{align}\label{eq:new-resolvent}
    \mathbbm{R}(\varepsilon)
    = \left[1 - \frac{\mathbbm{Q}}{\omega \mathbbm{M}}
    (\mathbbm{H}-\varepsilon)\right]^{-1}.
\end{align}
Since $\mathbbm{R}\mathbbm{P}^{(0)}\ket{\mathbbm{u}_\alpha}
= \ket{\mathbbm{u}_\alpha}$, multiplying both sides by
$\mathbbm{P}^{(0)}\mathbbm{H}$ yields
$\mathbbm{P}^{(0)}\mathbbm{H}\mathbbm{R}\mathbbm{P}^{(0)}
    \ket{\mathbbm{u}_\alpha}
    = \varepsilon_\alpha \mathbbm{P}^{(0)}\ket{\mathbbm{u}_\alpha}$,
and hence
\begin{align}\label{eq:new-H_eff}
    \mathbbm{H}_{\rm eff}(\varepsilon)
    = \mathbbm{P}^{(0)} \mathbbm{H} \mathbbm{R}(\varepsilon)
    \mathbbm{P}^{(0)}.
\end{align}

The explicit dependence on $\varepsilon$ can be eliminated using
\begin{align}
    \left[1 - \frac{\mathbbm{Q}}{\omega \mathbbm{M}}
    (\mathbbm{H}-\varepsilon)\right]
    \mathbbm{R}\mathbbm{P}^{(0)}\ket{\mathbbm{u}_\alpha}
    = \mathbbm{P}^{(0)}\ket{\mathbbm{u}_\alpha}.
\end{align}
By replacing $\varepsilon_\alpha \mathbbm{P}^{(0)}\ket{\mathbbm{u}_\alpha}$ by
$\mathbbm{H}_{\rm eff}\mathbbm{P}^{(0)}\ket{\mathbbm{u}_\alpha}$,
and then invoking the expression \eqref{eq:new-H_eff} for $\mathbbm{H}_{\rm eff}$, this leads to
\begin{align}
    \left[1 - \frac{\mathbbm{Q}}{\omega \mathbbm{M}}
    (1-\mathbbm{R}\mathbbm{P}^{(0)})\mathbbm{H}\right]
    \mathbbm{R}\mathbbm{P}^{(0)}\ket{\mathbbm{u}_\alpha}
    = \mathbbm{P}^{(0)}\ket{\mathbbm{u}_\alpha},
\end{align}
which implies the operator identity
\begin{align}\label{eq:resolvent-identity}
    \mathbbm{R}\mathbbm{P}^{(0)}
    = \mathbbm{P}^{(0)}
    + \frac{\mathbbm{Q}}{\omega \mathbbm{M}}
    (1-\mathbbm{R}\mathbbm{P}^{(0)})
    \mathbbm{H}\mathbbm{R}\mathbbm{P}^{(0)}.
\end{align}
Solving this equation for the resolvent and substituting the result into
Eq.~\eqref{eq:new-H_eff}, the effective Hamiltonian can be obtained. Energy-independent form  for $\mathbbm{R}$ enables a perturbative expansion of the effective Hamiltonian in powers of $1/\omega$ as 
\begin{align}
    \mathbbm{H}_{\rm eff} &= \mathbbm{P}^{(0)} \mathbbm{H} \mathbbm{R} \mathbbm{P}^{(0)} \nonumber\\
    &= \mathbbm{P}^{(0)} 
    \left(
        \mathbbm{H} + \mathbbm{H} \frac{\mathbbm{Q}}{\omega \mathbbm{M}} \mathbbm{H}
        + \mathbbm{H} \frac{\mathbbm{Q}}{\omega \mathbbm{M}} 
        \mathbbm{H} \frac{\mathbbm{Q}}{\omega \mathbbm{M}} \mathbbm{H} 
    \right.
    \nonumber\\
    & \left. \qquad\qquad\qquad
        - \mathbbm{H} \frac{\mathbbm{Q}}{\omega^2 \mathbbm{M}^2} \mathbbm{H}
        \mathbbm{P}^{(0)} \mathbbm{H}
        + \cdots
    \right)
    \mathbbm{P}^{(0)}.
\end{align}
Considering a single-frequency time-periodic Hamiltonian ${\cal H}(t)=H_0  + H_1 e^{-i\omega t} + H_{-1} e^{i\omega t}$, 
the effective Hamiltonian reads:
\begin{align}
    \mathbbm{H}_{\rm eff} 
    &= H_0 + \frac{1}{\omega} \left[ H_{-1}, H_1 \right] 
    \nonumber\\
    &+\frac{1}{\omega^2}  
    \left(
        H_{-1} \left[ H_0, H_1 \right] +
        H_{1} \left[ H_0, H_{-1} \right] \right)
    + \ldots,
\end{align}
in which we have used the fact that $\mathbbm{H}_{n,m}=H_{n-m}$.

\section{Dirac systems driven by structured light}
The Hamiltonian for a 2D Dirac system coupled to a mono-chromatic 
(single-frequency) field can be written as 
\begin{align}
    {\cal H}(t) =  v_F {\bm\sigma} \cdot \left[  {\bf p} -\frac{e}{  c} \widetilde{\bf A}({\bf r},t) \right] + m \sigma_z,
\end{align}
with $\tilde{\bf A}({\bf r},t) = {\bf A}({\bf r}) e^{-i\omega t} + {\bf A}^\ast({\bf r}) e^{i\omega t} $.
Therefore, the Fourier components of the Hamiltonian are given by
\begin{align}
    {H}_0 & =   v_F {\bm\sigma} \cdot  {\bf p} + m \sigma_z 
     = \begin{pmatrix}
        m &   v_F p_- \\
          v_F p_+ & -m    
    \end{pmatrix},
    \\
    {H}_1 & = -e\frac{v_F}{c}  {\bm\sigma} \cdot {\bf A}({\bf r}) 
    =  -e\frac{v_F}{ c} \begin{pmatrix}
        0 & A_- \\
        A_+ & 0
    \end{pmatrix},
    \\
    {H}_{-1} & = -e\frac{v_F}{ c} {\bm\sigma} \cdot {\bf A}^\ast({\bf r})
    =  -e\frac{v_F}{ c} \begin{pmatrix}
        0 & A_+^\ast \\
        A_-^\ast & 0
    \end{pmatrix},
\end{align}
with $p_\pm = p_x \pm i p_y$ and $A_\pm = A_{x}({\bf r})  \pm i A_{y}({\bf r}) $.
By plugging in the expressions for $H_0$, $H_1$, and $H_{-1}$,
the effective Hamiltonian up to the second order reads
\begin{align}
    \mathbbm{H}_{\rm eff} 
    &\approx    v_F {\bm\sigma} \cdot  {\bf p} + m \sigma_z \nonumber\\
    &+ \frac{e^2}{\omega}\left(\frac{ v_F}{c}\right)^2 
    \left[ 
        {\bm\sigma} \cdot {\bf A}^\ast({\bf r}) , 
        {\bm\sigma} \cdot {\bf A}({\bf r}) 
    \right]
    \nonumber\\
    &-\frac{e^2}{\omega^2}\left(\frac{ v_F}{c}\right)^2 
    \left[
        \left\{
            {\bm\sigma} \cdot {\bf A}^\ast({\bf r}) , 
            {\bm\sigma} \cdot {\bf A}({\bf r}) 
        \right\}
        \left( v_F {\bm\sigma} \cdot  {\bf p} + m \sigma_z \right)
    \right. \nonumber\\
    &\left. \qquad\qquad  -
        {\bm\sigma} \cdot {\bf A}^\ast({\bf r}) \:
        \left( v_F {\bm\sigma} \cdot  {\bf p} + m \sigma_z \right) \:
        {\bm\sigma} \cdot {\bf A}({\bf r})
        \right. \nonumber\\
    &\left. \qquad\qquad  -
        {\bm\sigma} \cdot {\bf A}({\bf r}) \:
        \left( v_F {\bm\sigma} \cdot  {\bf p} + m \sigma_z \right) \:
        {\bm\sigma} \cdot {\bf A}^\ast({\bf r})
        \right].
\end{align}  
We note that the effective Hamiltonian is position-dependent
and the momentum operator ${\bf p}$ should be treated as $-i\nabla$.

By evaluating the relevant commutators and anticommutators, and neglecting terms proportional to spatial derivatives of the vector potential (i.e., setting ${\bf p} [ {\bf A}({\bf r}) ] \sim 0$), we obtain an effective Hamiltonian of Dirac form with a (position-dependent) renormalized effective mass and velocities, as presented in the main text. We have also examined the contribution arising from the noncommutativity of momentum and position, which yields an effective gauge field
\begin{align}
\mathbfcal{A}^{\rm eff} =
2i v_F \frac{\xi}{\omega} \Re \big[ (\nabla \cdot {\bf A})\, {\bf A}^\ast
+
(\nabla\times {\bf A})_z\,
(\hat{\bf z}\times {\bf A}^\ast)
\big]
\end{align}
This additional term is, however, non-Hermitian and more precisely, anti-Hermitian. This is a known feature of BW perturbation theory, which is not guaranteed to preserve Hermiticity at intermediate stages. It is likewise known that, provided the perturbative expansion is kept consistent to the order of interest (eigenvalues and right eigenvectors are retained up to the same order $n$ to which the effective Hamiltonian has been expanded) such non-Hermitian contributions do not affect the results at that order, only appearing starting at order \cite{Aoki2016PRB}. We therefore safely drop this term without loss of physical content.

\section{Hamiltonian and Green's function in the Wigner representation}

To account for the spatially varying light-induced terms in the effective Hamiltonian, we employ the Wigner representation and perform a systematic expansion in spatial gradients. The retarded Green's function satisfies the Wigner-transformed Dyson equation \cite{Von-Eiff1992},
\begin{equation}
\left[E + i\eta - H_W(\mathbf{R}, \mathbf{k})\right] \star G_W(\mathbf{R}, \mathbf{k}; E) = \mathbb{I},
\end{equation} 
where $\mathbf{R}=(\mathbf{r}_1+\mathbf{r}_2)/2$, and 
we have Fourier-transformed the relative coordinate
$\mathbf{r}=\mathbf{r}_1-\mathbf{r}_2$. Also, $\star$ denotes the Moyal product that admits the gradient expansion \cite{ZHANG2020135197}
\begin{equation}
    A\star B=AB +\frac{i\hbar}{2}\{A,B\}_{\text{PB}} + \mathcal{O}(\hbar^2).
\end{equation} 
with phase-space Poisson bracket given as 
$
\left\{
A_W,B_W
\right\}_{\mathrm{PB}}
=
\nabla_{\mathbf{R}} A_W
\cdot
\nabla_{\mathbf{k}} B_W
-
\nabla_{\mathbf{k}} A_W
\cdot
\nabla_{\mathbf{R}} B_W$
where the order of the matrix products is retained. Similarly, the Wigner Green's function is expanded as $G_W
=
G_W^{(0)}
+
\hbar G_W^{(1)}
+
\mathcal{O}(\hbar^2)$
where
\begin{align}
    G^{(0)}_W &= (E+i\eta - H)^{-1}, \\
    G_W^{(1)} &= -\frac{i}{2}G_W^{(0)}\{H_W,G_W^{(0)}\}_{\text{PB}},
\end{align}
that correspond to the local adiabatic approximation and leading correction to the local approximation arising from spatial gradients of the effective Hamiltonian, respectively. By plugging these results for the Green's function in Eq. \eqref{eq:spin_density} spin density can be obtained.

\end{document}